\documentclass[lettersize,journal]{IEEEtran}
\usepackage{amsmath,amsfonts}
\usepackage{algorithmic}
\usepackage{algorithm}
\usepackage{array}
\usepackage{textcomp}
\usepackage{stfloats}
\usepackage{url}
\usepackage{verbatim}
\usepackage{graphicx}
\usepackage{cite}
\usepackage{subfigure}
\usepackage{caption}
\usepackage{subcaption}
\usepackage{color}
\usepackage{cuted}
\usepackage{etoolbox}

\begin{document}


\title{{RayProNet: A Neural Point Field Framework for Radio Propagation Modeling in 3D Environments}}

\author{Ge Cao and Zhen Peng,~\IEEEmembership{Senior Member,~IEEE}
\thanks{G. Cao and Z. Peng are with the Center for Computational Electromagnetics, Department of Electrical and Computer Engineering, University of Illinois at Urbana-Champaign, Urbana, IL 61801 USA (e-mail: gecao2@illinois.edu;  zvpeng@illinois.edu).}}

\newcommand{\ZP}[1]{{\color{black}#1}}
\newcommand{\GC}[1]{{\color{black}#1}}

\maketitle

\hspace*{\fill}

\hspace*{\fill}

\hspace*{\fill}

\hspace*{\fill}

\hspace*{\fill}

\hspace*{\fill}

\begin{strip}
\begin{minipage}{\textwidth}\centering
\vspace{-100pt}
\includegraphics[width=0.995\textwidth]{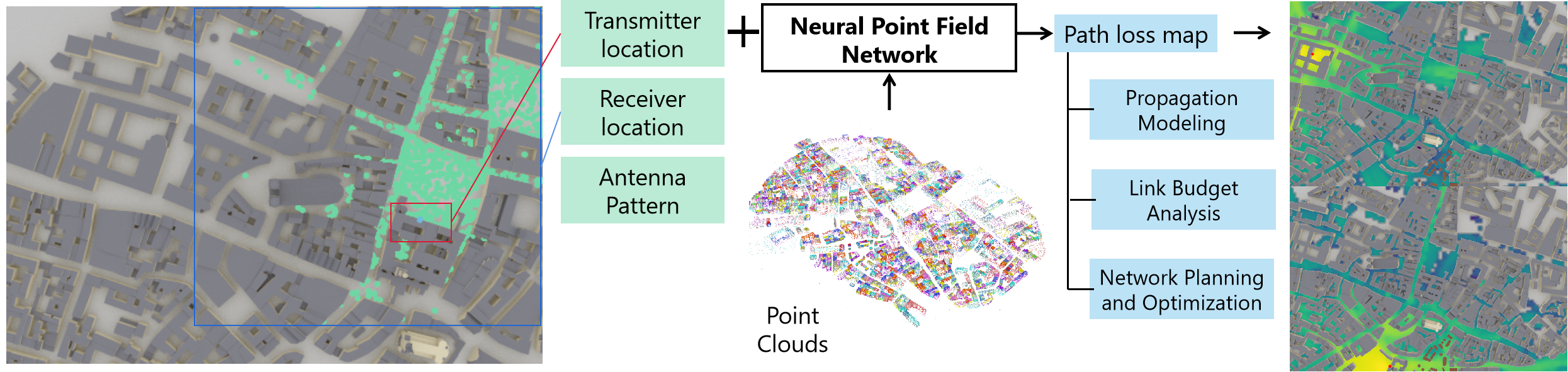}
\captionof{figure}{
{The schematic illustrates the input, output, and application of our proposed neural point field network framework for predicting wireless radio channel properties in large-scale environments.}
}
\label{figurelabel}
\end{minipage}
\end{strip}

\begin{abstract}
The radio wave propagation channel is central to the performance of wireless communication systems. 
In this paper, we introduce a novel machine learning-empowered methodology for wireless channel modeling.
The key ingredients include a point-cloud-based neural network and a Spherical Harmonics encoder with light probes. 
Our approach offers several significant advantages, including the flexibility to adjust antenna radiation patterns and transmitter/receiver locations, the capability to predict radio power maps, and the scalability of large-scale wireless scenes. As a result, it lays the groundwork for an end-to-end pipeline for network planning and deployment optimization. The proposed work is validated in various outdoor and indoor radio environments.

\end{abstract}

\IEEEpeerreviewmaketitle

\section{Introduction}
{Understanding and accurately modeling the characteristics of the propagation channel are essential for the design, deployment, and optimization of wireless communication networks \cite{2009radio, Mondal20153DCM, Tapan2018}. 
Although Maxwell's Equations govern the fundamental physics of wireless information transmission, obtaining full-wave solutions in large-scale environments is typically challenging and time-consuming \cite{865237, Tsang_2004, 1504967, Sevgi_2007_Hybrid,  Sarris_FDTD_Fading, 8485766}. 
Ray tracing-based simulators are commonly employed for modeling wireless channel properties \cite{Aguado_2000, Sarkar_Ray_2001, 901882, Raytracing_2015, Raytracing_tutorial_2019}. In the ray tracing process,  electromagnetic (EM) rays are uniformly launched from the transmitter antenna, undergoing reflections, transmissions, and diffractions with various buildings and floors, ultimately reaching the receiver locations. These ray paths and interactions yield valuable wireless channel information such as channel gain, channel transfer function, and channel impulse response.} 

{While ray tracing has been a popular tool in wireless channel modeling, its computational complexity escalates with the number of ray-object interactions. 
Moreover, in wireless deployment and planning scenarios, frequent modifications to transmitter/receiver locations are common. Typically, a new ray tracing simulation is required for each configuration change. This exhibits a noticeable gap between the simulation time of ray-tracing simulators and the rapid time-to-solution demand of wireless network design and optimization. 
To address these needs, neural network-based forward surrogate models emerge as an attractive solution \cite{Ray-Launching-Neural-2014, 8740286, Costas_2022, Seretis_2023}. Neural networks generally offer faster runtime compared to ray tracing algorithms, and their accuracy can be enhanced by refining the training dataset rather than increasing runtime. 
}

{The objective of this paper is to develop a neural network surrogate capable of predicting wireless channel properties across large-scale environments. The overview of the proposed framework is given in Fig. \ref{figurelabel}.
In our methodology, we train the neural surrogate using ray-tracing solutions corresponding to a finite set of transmitting locations within a specific radio environment. Once trained, the neural surrogate leverages its understanding of EM propagation physics to predict EM wave propagation for new transmitter/receiver locations and different antenna radiation patterns.
This research emphasizes two key features: (1) the neural surrogate's functionality to predict the spatial distribution of radiated power (i.e., the radio coverage or path loss map), and (2) its effective generalization to large-scale scenes in both outdoor and indoor environments.
}

{In the realm of neural surrogate development for radio wave propagation, the learning of scene representations is an aspect that has received limited attention in previous works. 
Many existing approaches primarily focus on 2D image tasks, typically from a bird's-eye view, and lack the incorporation of geometry information as input \cite{hehn2023transformer, lee2023pmnet, DeepRay2022}. Another recent study \cite{WiNert2023} focuses on explicitly learning the meshed geometries, thereby limiting its generalizability to large outdoor scenes.
In contrast, our proposed work offers a fresh perspective on neural scene representation.
The 3D propagation environment (wireless scene) is rendered using point clouds, a representation well-known for its adaptability and intuitive scalability to large-scale scenes \cite{ost2022pointlightfields}. 
}

{Moreover, we introduce the Neural Point Field framework to implicitly embed wireless channel state information into light probes \cite{lightprobe2008}. 
Each light probe encapsulates EM ray properties, which are interpolated using a Spherical Harmonics {encoder and decoder \cite{Renerf2023}}. This facilitates the extraction of propagation information from queries in different ray directions.
Conceptually, {these light probes are designed to capture the site-specific EM ray propagation physics. Receivers can seamlessly extract path tracing and ray propagation from these probes, streamlining the process and enhancing overall efficiency.}
}

{Compared to existing neural ray tracing methods in the literature, the proposed work excels in scalability and flexibility, accommodating diverse levels of geometry complexity while maintaining high-quality channel prediction. 
We validate our proposed pipeline across small indoor, medium outdoor, and large city scenes. The results demonstrate the efficacy of our approach in predicting wireless channel properties across various scales of scenes.}

\section{Related Works}

In this section, we discuss related works from both the machine learning (ML) and wireless communication communities. Given the resemblance between rendering and wireless channel modeling algorithms, we particularly emphasize studies in neural rendering and {computer} graphics within the deep learning field. Additionally, since our pipeline design necessitates an implicit representation of geometry, we also introduce relevant works on geometry in neural networks.

\textbf{Neural Rendering:} {The ray tracing algorithm is widely used in the rendering process in 3D computer graphics}. {Leveraging this foundational understanding, our research explores valuable insights from advancements in neural rendering, enriching our approach to wireless channel modeling.}
Recently, advancements in 3D scene representation using neural networks have showcased their ability to render scenes quickly and flexibly. In these approaches, the radiance field is embedded within neural networks, such as Multi-Layer Perceptrons (MLPs), or at a higher level, within the volume space. This implies that the lighting information is typically fixed and cannot be modified. Despite the complexity of light sources in the rendering process \cite{Haocheng2023}, several works have achieved relighting techniques \cite{NeRFactor2021, EyeNeRF2022, Renerf2023}.

Since the publication of Kerbl et al.'s work on 3D Gaussian Splatting \cite{3Dgaussians2023}, this new neural rendering technique has garnered significant attention. A Gaussian kernel is applied and learned to represent scene geometries in the format of point clouds. Subsequently, several related works have emerged, including research on relighting \cite{Relightable3DGaussian2023} and the reconstruction of human avatars \cite{liu2023animatable}.

{Before the development of 3D Gaussian Splatting, a strategy known as Neural Point Light Fields (NeuralPointLF) was introduced, demonstrating the potential of point cloud formats in the domain of neural rendering \cite{ost2022pointlightfields}. The distinction between NeuralPointLF and 3D Gaussian Splatting lies in the fact that NeuralPointLF does not necessitate a rasterization process in the pipeline. Since ray-tracing simulations in wireless channel modeling also do not require rasterization, our network draws inspiration from NeuralPointLF and incorporates attention techniques into the framework \cite{Vaswani2017attention}. 
{Furthermore, as NeuralPointLF lacks a relighting process, our pipeline incorporates relighting into its design. This addition addresses scenarios involving changing antenna locations or radiation patterns.
}
}

\textbf{Geometry Representation:} The representation format of 3D geometry is crucial for all ML tasks involving three-dimensional data. The most common method for representing geometry is through mesh triangles, consisting of a set of vertices ($\mathbb{V}$), edges ($\mathbb{E}$), and faces ($\mathbb{F}$). 
{While meshed geometries are widely utilized in computational science and engineering, their utilization in deep learning is limited due to the non-differentiability of triangle face indices.}
 Although some researchers have attempted to apply statistical methods to make mesh triangles differentiable \cite{DiffRayTracing2018, Mitsuba3}, these strategies are still computationally intensive for neural networks.

{Point clouds have emerged as a preferred geometry representation format in neural network-based research. This representation is utilized across various tasks, including 3D surface reconstruction \cite{DeepSDF2019, DeepPointCloudeReconstruction2021}, geometry denoising \cite{PointProNets2018, pointcleannet2020, Luo_2021_ICCV}, and geometry completion \cite{wen2021pmp, xiang2021snowflakenet}. 
Leveraging the differentiability of point clouds, our work adopts the \emph{PointNet} \cite{qi2016pointnet} architecture for geometry representation. While mesh triangles and point clouds are prevalent, other representation formats exist, such as the multi-view model \cite{MutiView2018, rotationnet2018} and surface random walk \cite{AttWalk2022}.}

\textbf{Neural Radio Channel Modelling:} Physically-based simulation guided by neural networks is gaining popularity across various {scientific} domains, including fluid dynamics \cite{xie2018tempoGAN, StreamGuidedFluid2021}, soft body dynamics \cite{diffpd, DiffAqua}, and electrodynamics \cite{ZP2022, ZP2024}, etc. In the field of applied and computational electromagnetics, several approaches leveraging neural networks have been proposed \cite{Xudong2019, Li2019, Yao2020, Xu2020, Yao2022, Ge2022, Guo2022, JiefuChen2020, Maokun_2022, JiefuChen2022, Abdulkadir_2023, Maokun_2023, JiefuChen2023, Shutong_2024}. Many of these neural surrogates aim to learn the scattering process involving obstacles in free space. 
{Given that wireless channel properties are governed by the propagation and scattering of EM waves, our work shares objectives related to those of these approaches}. 
The emphasis of this work is to expand the application domain to encompass more complex scenarios, specifically extending into 3D environments featuring intricate obstacles like buildings.

Until now, there has been limited attention given to the task of wireless channel modeling in complex 3D environments. A recent work addressing this task is \emph{WINERT} \cite{WiNert2023}. In their approach, a complete ray tracing process is implemented, with a focus on learning the propagation properties (reflection, transmission, diffraction) of buildings. However, they did not implement the ray-triangle intersection process as differentiable, citing its non-differentiability. Furthermore, their pipeline is not suitable for handling large-scale and complicated scene geometries.

{Several other works have also aimed to develop neural surrogates for predicting path loss map information. Nevertheless, most of these works focus on 2D tasks that do not explicitly require geometry representation. Instead, they rely on 2D bird's-eye-view images (heatmaps) for training \cite{hehn2023transformer, lee2023pmnet, DeepRay2022}. While this format simplifies the learning process and results in a faster pipeline, it may encounter difficulties in effectively capturing the complexities of 3D scenes in an end-to-end manner.
}

\section{Neural Point Field for Wireless Channel}

\begin{figure*}[!t]
\centering
\includegraphics[width=0.995\textwidth]{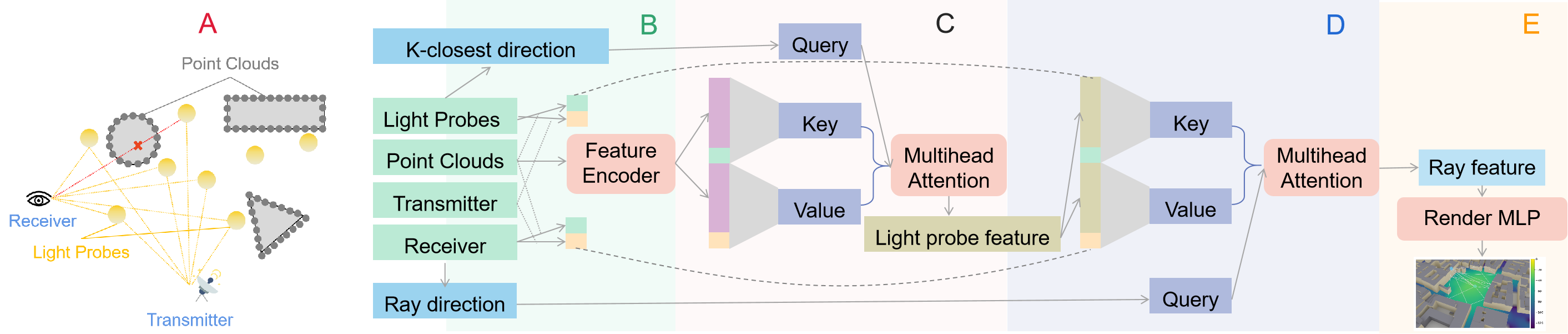}
\caption{{RayProNet: a neural point field framework for wireless channel modeling pipeline. (The symbols A - E represent the subsections in Section III.)}
}
\label{fig: pipeline}
\end{figure*}

{The proposed work aims to investigate a neural point field network to simulate the ray tracing process between transmitters and receivers within complex wireless scenes. 
At its core, this method relies on three fundamental elements: leveraging point clouds for the representation of geometric structures, integrating light probes to capture path tracing and ray propagation information, and utilizing spherical harmonic functions for the extraction of field data.
 An overview of the pipeline is illustrated in Fig. \ref{fig: pipeline}, which we henceforth refer to as RayProNet. The detailed technical ingredients and underlying rationales are provided below.
}

\subsection{Data Preparation}

\begin{figure}
\centering
\includegraphics[width=0.46\textwidth]{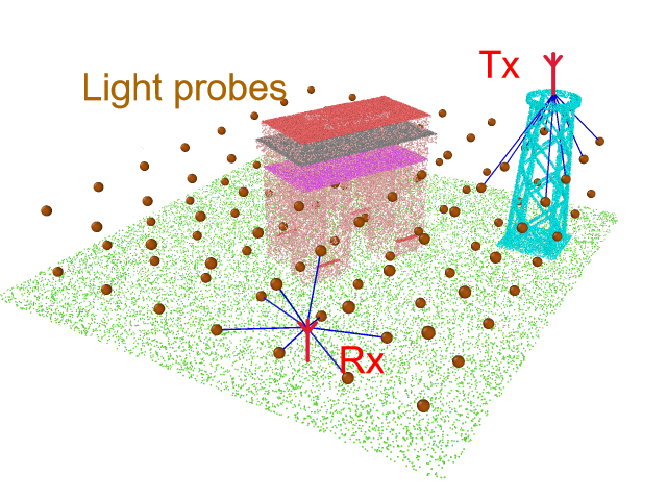}
\caption{{\textbf{Identifying n-nearest light probes}: Each receiver locates its $n$ nearest light probes and retrieves radiance information from them.}}
\label{fig: light probes - rx}
\end{figure}

\begin{figure}
\centering
\includegraphics[width=0.46\textwidth]{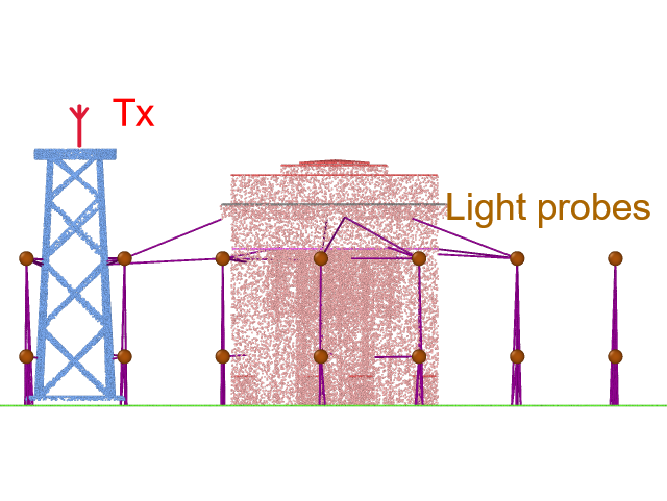}
\caption{{\textbf{Identifying K-nearest points}: Each light probe finds its $K$ closest points and encodes occlusion information.}}
\label{fig: pts - light probes}
\end{figure}

{
The RayProNet pipeline relies on two primary inputs: the locations of receivers and transmitters, alongside the 3D geometry of the environment, which is initially transformed into point clouds as the default format for representation. Point clouds offer an efficient means of encoding complex geometric features like obstacles, buildings, and terrain by sampling points in space to capture the characteristics of interacting objects within the environment. This approach allows for the effective encoding of interacting objects, with a particular emphasis on learning geometric features.

In addition, light probes are uniformly placed throughout the scene, capturing the propagation behavior of EM rays through space. 
Their integration into the pipeline allows the model to acquire essential insights into ray paths, reflections, and diffractions, thereby enhancing the accuracy and efficiency of the learning process.
Light probes play a crucial role in encoding propagation information, particularly in environments characterized by sparse geometric structures, as elaborated in Section III.C. 
The data preparation stage proceeds as follows:
}
{
\begin{itemize}
\item \textbf{Transmitter Setup:} Initially, we define the locations of transmitters and configure their antenna patterns. This process ensures an accurate representation of transmitter characteristics in the simulation.
\item \textbf{Receiver Setup:} Similarly, we specify the locations of receivers and configure their antenna patterns to accurately simulate receiver behavior in the wireless environment.
\item \textbf{Identify $n$ Nearest Light Probes:} For each receiver, we identify the $n$ nearest light probes and record ray directions, enabling the collection of electromagnetic field information from the surroundings (Fig. \ref{fig: light probes - rx}).
\item \textbf{Identify $K$ Nearest {Points}:} Next, we determine the $K$ nearest {points} for each light probe and record this as a $K$-closest direction attachment. This enables us to capture detailed geometric information about the scene (Fig. \ref{fig: pts - light probes}).
\end{itemize}
The parameters $n$ and $K$ serve as hyperparameters that offer flexibility for customization based on scene complexity and application-specific consideration, allowing for tailored adjustments to the pipeline.  For example, applications requiring highly fidelity predictions or precise localization may benefit from larger values of $n$ and $K$.
}

\subsection{Point Cloud Feature Embedding}

Given our primary focus on wireless channel modeling rather than rendering, employing a multi-view model presents challenges due to the absence of a specific look-at direction in our task. Therefore, we adopt the PointNet model \cite{qi2016pointnet} for its effectiveness and robustness in learning various features of point clouds. 
PointNet is originally proposed for 3D recognition tasks such as object classification, part segmentation, and semantic segmentation. Unlike traditional convolutional neural networks (CNNs) that operate on grid-like data and images, PointNet can directly process point clouds without requiring any intermediate representation like voxelization.
In our implementation, we begin by normalizing all scene point clouds to the range $[-1, 1]$. We then utilize PointNet to generate a feature matrix $\boldsymbol{l}_{j,k} \in \mathbb{R}^{n_{p} \times 128}$, where $n_{p}$ is the total number of point clouds.
Afterwards, we split this matrix into $\boldsymbol{l}_{j,k1} \in \mathbb{R}^{n_{p} \times 64}$ and $\boldsymbol{l}_{j,k2} \in \mathbb{R}^{n_{p} \times 64}$ for use in subsequent interpolation steps.

\subsection{Path Tracing with Light Probes and Point Clouds}

Integrating light probes into our pipeline stands as an important contribution to our pipeline. 
It represents a strategic solution to address the unique challenges encountered in wireless ray propagation scenarios. In environments characterized by sparse geometric structures, such as open landscapes or urban settings with tall buildings, a straightforward implementation of neural ray tracing may encounter limitations. 
When rays emitted from antennas fail to intersect with nearby point clouds, one has to extrapolate their trajectories into unobstructed space.

To mitigate potential inaccuracies arising from the absence of precise ray directions, the proposed work draws inspiration from the concept of light probes in computer graphics. Light probes serve as essential tools for capturing and simulating realistic lighting effects within virtual environments. These probes act as virtual cameras that record light information from different directions, allowing for the creation of dynamic and immersive lighting scenarios.

In our pipeline, we place a set of light probes (much fewer than the number of point clouds)  throughout the scene. Each light probe serves as a virtual observation point, capturing and encoding surrounding ray propagation information. This encoded data allows nearby receivers to easily decode it using the ray direction and distance as queries.  Essentially, individual light probes serve as a neural surrogate for baking the propagation information within their nearby space.

Moreover, the introduction of light probes enables the extraction of EM field information from the embedded features of point clouds. Specifically, we select the $K$ closest point clouds for each light probe. Analogous to the relighting task in neural rendering, our re-transmitting task requires considering the contributions from the transmitter locations. As a result, for each light probe, a total of $K + 1$ points, comprising both point clouds and transmitters, are chosen for providing the information of transmitter signal propagation and the occlusion of buildings. This selection establishes a physical correspondence, where there is a Line of Sight (LOS) contribution from the transmitter (akin to direct illumination in rendering) and $K$ contributions (resembling wave physics of reflection, diffraction, and scattering) from point clouds.

\begin{figure}
\centering
\includegraphics[width=0.48\textwidth]{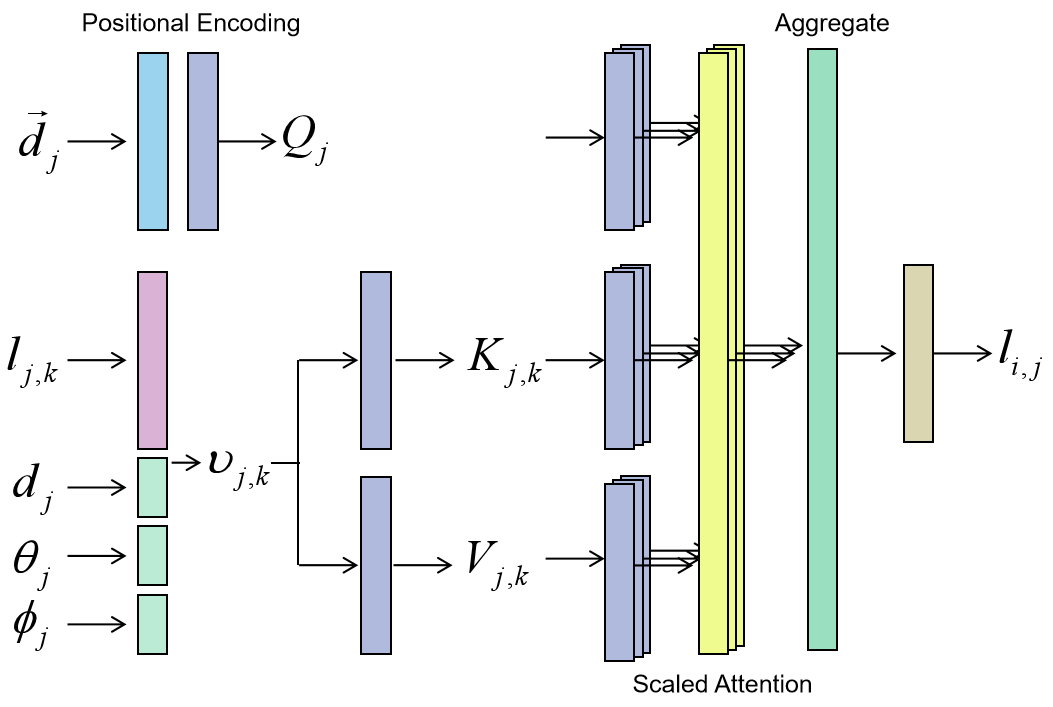}
\caption{\textbf{Multi-head attention}: In Section III.C, a multi-head attention module is employed to aggregate the point cloud feature vector $l_{j,k}$ along with the K-closest direction $j$. This process generates a light probe feature. The attention module described in Section III.D follows a similar structure.}
\label{fig: attention}
\end{figure}

In our design, an attention technique is employed for this extraction process (Fig. \ref{fig: attention}), guided by the location information between light probes and transmitter (distance $d_{t}$, elevation $\theta_{t}$ and azimuth $\phi_{t}$), and the information between light probes and point clouds (distance $d_{j}$, elevation $\theta_{j}$ and azimuth $\phi_{j}$). Subsequently, we combine them with our previous embedded feature as $\boldsymbol{K}_{j,k} \in \mathbb{R}^{K \times 67}$ and $\boldsymbol{V}_{j,k} \in \mathbb{R}^{K \times 67}$.

\begin{equation}
\begin{aligned}
\centering
\begin{cases}
\boldsymbol{K}_{j,k} = \boldsymbol{l}_{j,k1} \oplus \{d_{j}, \theta_{j}, \phi_{j} \} \\
\boldsymbol{V}_{j,k} = \boldsymbol{l}_{j,k2} \oplus \{d_{t}, \theta_{t}, \phi_{t} \} \\
\end{cases}
\label{Equation: Key-Value From point clouds to light probes}
\end{aligned}
\end{equation}

By adding a query into the pipeline, which is instructed by $K + 1$ closest direction $\boldsymbol{d}_{j}$ from light probes to point clouds and transmitter, EM field profiles are effectively baked into our light probes. These closest directions are shot from light probes directly to point clouds, and positionally encoded by encoder $\boldsymbol{F}_{\theta_{Q}}$ (\ref{Equation: Query From point clouds to light probes}).

\begin{equation}
\begin{aligned}
\centering
\boldsymbol{Q}_{j} = \boldsymbol{F}_{\theta_{Q}}(\boldsymbol{d}_{j})
\label{Equation: Query From point clouds to light probes}
\end{aligned}
\end{equation}

Subsequently, we apply multi-head attention to learn the power information of light probes: Given key-value pair $(\boldsymbol{k}_{j,k}, \boldsymbol{V}_{j,k})$, the task is to predict a weight corresponding to the query ray vector $\boldsymbol{Q}_{j}$. The output weight is then encoded into point cloud feature vector $\boldsymbol{l}_{i,j} \in \mathbb{R}^{n \times 128}$.

\begin{equation}
\begin{aligned}
\centering
\boldsymbol{l}_{i,j} = \boldsymbol{F}_{\theta_{atten}}(\boldsymbol{K}_{j,k}, \boldsymbol{V}_{j,k}, \boldsymbol{Q}_{j})
\label{Equation: Attention From point clouds to light probes}
\end{aligned}
\end{equation}

\subsection{Receivers: Unveiling Ray Physics from Light Probes}
Once the EM propagation information has been baked into light probes, the next step is to determine a format of interpolation for storing this data. Similar to the previous section, we employ another attention technique when receivers extract EM power from light probes. In this process, we select the $n$ closest light probes and generate a ray for each of them. The direction of these rays serves as the query for our attention block.
 Thereby, we design two instructions (key and value) with the combination of three variables: distance $d_{i}$, elevation $\theta_{i}$, and azimuth $\phi_{i}$. They are between point clouds and receivers (key), transmitters and receivers (value). This section is very similar to the previous part.

 \begin{equation}
\begin{aligned}
\centering
\begin{cases}
\boldsymbol{K}_{i,j} = \boldsymbol{l}_{i,j1} \oplus \{d_{i}, \theta_{i}, \phi_{i} \} \\
\boldsymbol{V}_{i,j} = \boldsymbol{l}_{i,j2} \oplus \{d_{t}, \theta_{t}, \phi_{t} \} \\
\end{cases}
\label{Equation: Key-Value From rx to point clouds}
\end{aligned}
\end{equation}

{Upon receiving a key-value pair, we encode $n$ ray directions $\boldsymbol{d}_{i}$, which is shot from receivers to light probes. Following positional encoding, a query vector $\boldsymbol{Q}_{i}$ is generated, which is then applied to another Multi-head attention neural block for feature extraction. The resulting output is a ray feature vector $\boldsymbol{l}_{i} \in \mathbb{R}^{n+1}$, where $n$ represents the number of rays. Notably, the inclusion of LoS necessitates the addition of another receiver-transmitter ray into our pipeline, thereby augmenting the final feature count to $n + 1$ instead of $n$.

\begin{equation}
\begin{aligned}
\centering
\boldsymbol{Q}_{i} = \boldsymbol{F}_{\theta_{Q}}(\boldsymbol{d}_{i})
\label{Equation: Query From rx to point clouds}
\end{aligned}
\end{equation}

\begin{equation}
\begin{aligned}
\centering
\boldsymbol{l}_{i} = \boldsymbol{F}_{\theta_{atten}}(\boldsymbol{K}_{i,j}, \boldsymbol{V}_{i,j}, \boldsymbol{Q}_{i})
\label{Equation: Attention From rx to point clouds}
\end{aligned}
\end{equation}

\subsection{Spherical Harmonics-based Decoding of Ray Features}
After decoding the ray features from our attention neural blocks, we represent this information as a set of spherical harmonics coefficients. 
Spherical harmonics are special functions defined on the surface of a sphere, widely utilized in various fields such as atomic and molecular physics, quantum mechanics, and computer graphics. These functions constitute an orthogonal and complete set of basis functions, particularly renowned for their utility in encoding or decoding directional information.
A visualization of 3-order Spherical Harmonics is shown in Fig. \ref{fig: SH}, where a higher order suggests enhanced performance in restoring higher frequency and directional information.

In the previous subsection, a ray feature $\boldsymbol{l}_{i} \in \mathbb{R}^{n+1}$ is provided. Here, we employ an 8-layer multi-layer perceptron (MLP) with 256 channels. The output is a spherical harmonics interpolation coefficient $\boldsymbol{c}_{i} \in \mathbb{R}^{(n + 1) \times n_c}$, where $n_{c}$ is the output channel, typically set as the interpolation degree.

\begin{equation}
\begin{aligned}
\centering
\boldsymbol{c}_{i} = \boldsymbol{F}_{\theta_{MLP}}(\boldsymbol{l}_{i})
\label{Equation: Render MLP 1}
\end{aligned}
\end{equation}

Subsequently, we divide the output coefficient $\boldsymbol{c}_{i} \in \mathbb{R}^{(n + 1) \times n_c}$ into $\boldsymbol{c}_{i1} \in \mathbb{R}^{n \times n_c}$ and $\boldsymbol{c}_{i2} \in \mathbb{R}^{n_c}$. Finally, a Spherical Harmonics decoder is applied.

\begin{equation}
\begin{aligned}
\centering
\boldsymbol{o}_{i} = \boldsymbol{c}_{i1}^{T}SH(\boldsymbol{d}_{i}) + \boldsymbol{c}_{i2}
\label{Equation: Render MLP 2}
\end{aligned}
\end{equation}

\begin{figure}
\centering
\includegraphics[width=0.48\textwidth]{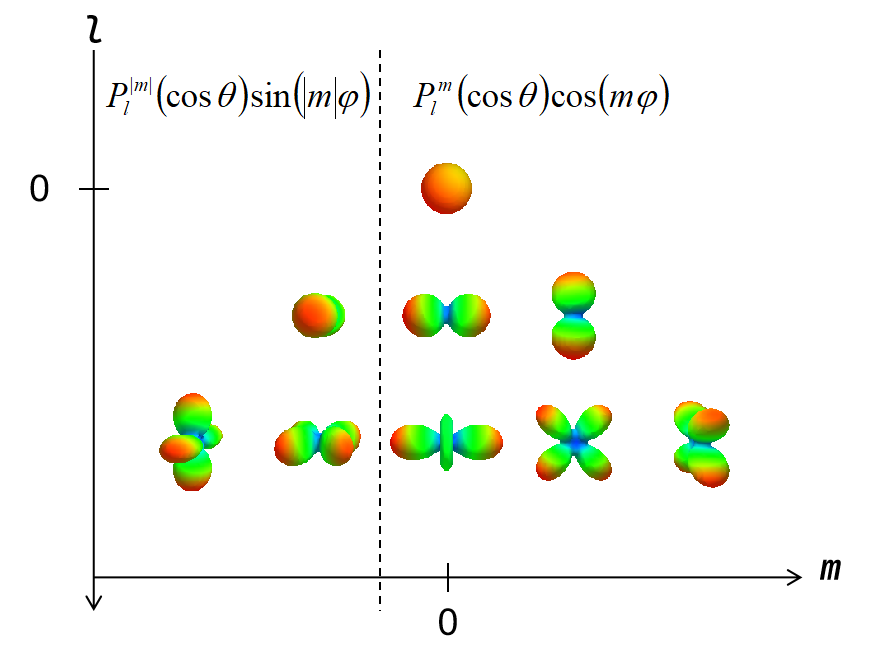}
\caption{\textbf{Spherical Harmonics}: This figure visualizes $3^\textrm{rd}$-order Spherical Harmonics, {whose solution is a multiple of the associated Legendre polynomial $P_{l}^{|m|}$ with input of azimuth $\varphi$ and elevation $\theta$}. In our methodology, Spherical Harmonics, which takes the ray direction as input, are employed for radiance encoding.
}
\label{fig: SH}
\end{figure}

\section{Numerical Experiments}

\begin{figure*}[!ht]
\centering
\subfigure{
    \rotatebox{90}{\scriptsize{~~~~~~~~~~GT}}
    \begin{minipage}[t]{0.22\textwidth}
        \includegraphics[width=\linewidth]{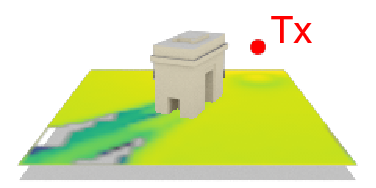}
    \end{minipage}
    \begin{minipage}[t]{0.22\textwidth}
        \includegraphics[width=\linewidth]{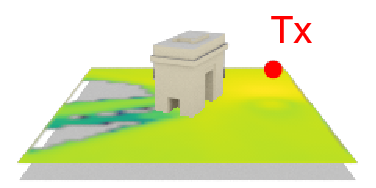}
    \end{minipage}
}
\subfigure{
    \begin{minipage}[t]{0.22\textwidth}
        \includegraphics[width=\linewidth]{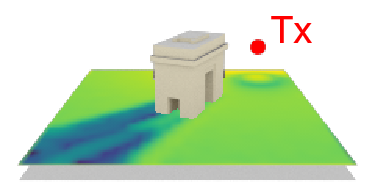}
    \end{minipage}
    \begin{minipage}[t]{0.22\textwidth}
        \includegraphics[width=\linewidth]{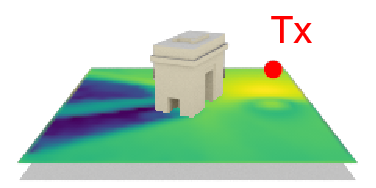}
    \end{minipage}
}
\\
\setcounter{subfigure}{0}
\subfigure[w/o diffraction]{
    \rotatebox{90}{\scriptsize{~~~~~~~~Ours}}
    \begin{minipage}[t]{0.22\textwidth}
        \includegraphics[width=\linewidth]{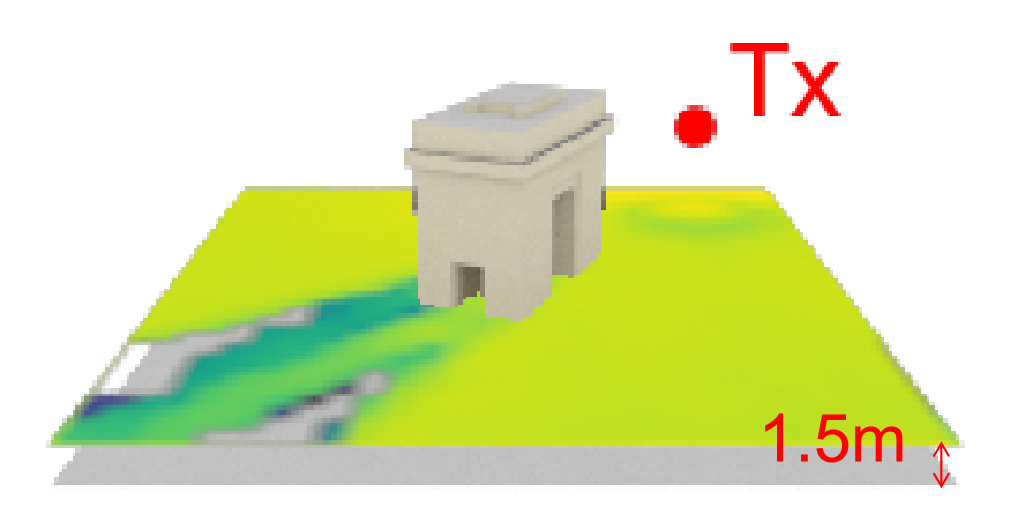}
    \end{minipage}
    \begin{minipage}[t]{0.22\textwidth}
        \includegraphics[width=\linewidth]{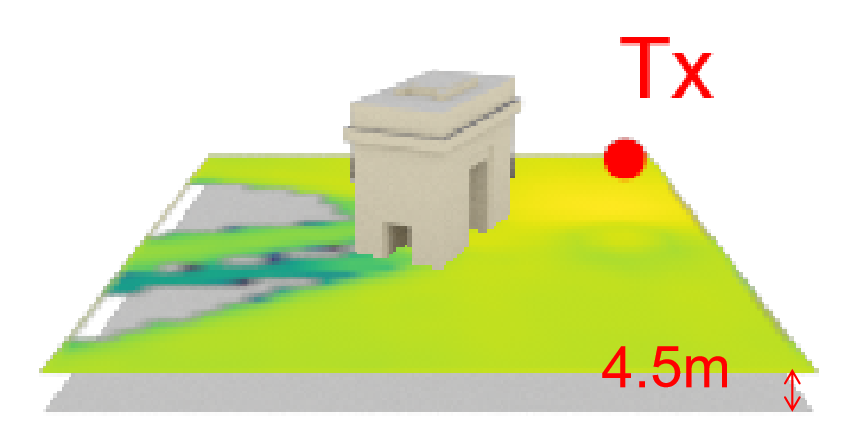}
    \end{minipage}
}
\subfigure[w/ diffraction]{
    \begin{minipage}[t]{0.22\textwidth}
        \includegraphics[width=\linewidth]{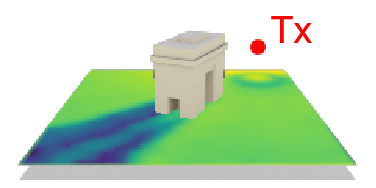}
    \end{minipage}
    \begin{minipage}[t]{0.22\textwidth}
        \includegraphics[width=\linewidth]{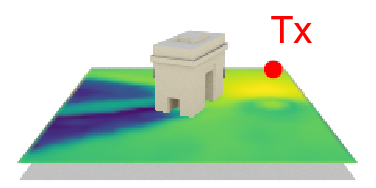}
    \end{minipage}
}
\caption{\textbf{Learning EM Physics with an isolated building:} This figure illustrates the application of our pipeline using a scene with a single isolated building at the center. The results indicate that our framework accurately captures and understands the principles of EM physics. (GT: Ground Truth using ray-tracing)}
\label{fig: Evaluation EM physics understanding}
\end{figure*}

{In this section, we quantitatively validate our proposed method for predicting radio propagation characteristics across various wireless environments. The section is divided into three parts. Part \textit{A} details the experimental setup, data collection, and the training process employed. Part \textit{B} focuses on the validation and verification procedures. Finally, Part \textit{C} demonstrates and evaluates the proposed methodology in large-scale, 3D wireless scenarios.}

\subsection{Experimental Setup and Training}

{Our evaluation aims to assess the effectiveness of the proposed model in accurately predicting signal strength and coverage within wireless environments. We focus on two key outcomes: path loss maps and received signal strength at designated locations. Path loss maps depict the attenuation of electromagnetic signals as they propagate through the wireless scenes, offering valuable insights into coverage areas. Meanwhile, the received signal strength at specific locations provides crucial information for tasks such as localization and connectivity assessment. Both outcomes are essential for network planning and optimization, informing decisions regarding antenna placement and transmission power levels to optimize network performance and reliability.}

\subsubsection{Data Collection}
{
Our datasets are generated using an open-sourced ray-tracing simulator: \emph{Sionna} \cite{sionna}. We generate our dataset in various scales of scenes such as: \emph{wiindoor} (small indoor room scene), \emph{etoile} (Medium city block scene), and Munich (large urban city scene).

As described in Table \ref{Table: dataset}, the dataset comprises $175 \sim 375$ transmitter locations and approximately $1,920 \sim 35,816$ uniformly sampled receiver locations for each scene. About $85\%$ of them is used for training, with the remaining $15\%$ reserved for validation. The operating frequency is $2.14 \rm{GHz}$. After training, the model serves as a neural surrogate for wireless channel prediction.
}

\begin{table*}
\begin{center}
\caption{\textbf{Data collection}: We validate our methodology across three different scene scales. This table provides the configuration details of our dataset.}
\label{Table: dataset}
\begin{tabular}{|c|c|c|c|c|}
 \hline
  \textbf{Training Dataset} & \emph{wiindoor} & \emph{etoile center} & \emph{etoile} & \emph{munich} \\
  \hline
 \textbf{Scale} & indoor room & isolated building & city blocks & urban city \\
 \hline
 \textbf{Covered area} & $10 \times 10 \;\rm{m}^2$ & $150 \times 160 \;\rm{m}^2$ & $853 \times 676 \;\rm{m}^2$ & $1475 \times 1205 \;\rm{m}^2$ \\
 \hline
 \textbf{Transmitters} & $175$ & $375$ & $175$ & $175$\\
 \hline
 \textbf{Receivers} & $100 \times 100 \times 2$ & $30 \times 32 \times 2$ & $86 \times 68 \times 2$ & $148 \times 121 \times 2$\\
 \hline
 \textbf{Antenna patterns} & \multicolumn{4}{c|}{$4$}\\
 \hline
\end{tabular}
\end{center}
\end{table*}

\subsubsection{Training setups}
In our experiments, $n$ rays are initially launched from receivers to find $n$ nearest light probes. Each light probe is attached to $K$ different point clouds. The hyperparameters $n$ and $K$ are selected as $8$.

Our model is trained with a batch size of $1000$ and a learning rate of $0.0001$. We train the model for $500$ epochs, which typically takes between $1.5 \sim 10$ hours in a GPU environment using an NVIDIA GeForce RTX 3090 Ti. We utilize the Adam optimizer \cite{Adam2015} and the mean square error (MSE) loss function for received power optimization.

\subsubsection{Evaluation Metric}

The evaluation metric serves as a quantitative measure to assess the performance of the proposed method in predicting radio path loss maps. It quantifies the accuracy of the predictions by comparing them to ground truth (GT) data or measurements. {The specific evaluation metrics that will be used in our numerical experiments include Mean Square Error (MSE), and Peak Signal-to-Noise Ratio (PSNR).}

\begin{equation}
\begin{aligned}
\centering
\begin{cases}
\rm{MSE} = \sum_{i=0}^{N} (o_{i} - o_{gt})^{2} \\
\rm{PSNR} = 20 \rm{log}_{10}({\rm{max}{(o_{i})}} / {\sqrt{MSE}}) \\
\end{cases}
\label{Equation: MSE and PSNR}
\end{aligned}
\end{equation}

\subsection{Validation and Verification}

\subsubsection{Assessment in Learning EM Propagation Physics}
{
In this subsection, we evaluate our framework's ability to model EM propagation physics. Our objective is to determine how effectively the neural network learns and understands key EM principles such as reflection, transmission, and diffraction, particularly in the context of various building structures. To assess this, we train the neural network using scenes with a single isolated building at the center. This setup allows us to isolate and analyze the model's performance in understanding EM propagation in the presence of architectural elements. 
We provide two separate datasets for training: one including diffraction effects and the other without. 
The validation results, shown in  Fig. \ref{fig: Evaluation EM physics understanding}, demonstrate the model's proficiency in accurately capturing essential features of EM propagation physics. Moreover, enhancements in prediction accuracy can be achieved through the refinement of the training dataset.
}

\subsubsection{Comparison to Other Neural Surrogates}
{While our primary focus is on 3D end-to-end channel power prediction, the scarcity of open-source 3D-based neural surrogates led us to evaluate our model against a 2D-based neural surrogate PMNet \cite{lee2023pmnet} and a standard Multi-Layer Perceptron (MLP) model. The MLP network is designed with four hidden layers of sizes 64, 64, 32, and 64, using leaky ReLU as the activation function.

Figure \ref{fig: Comparison with others} and Table \ref{Table: comparison with others} present the visualization and quantitative comparison. The results indicate that our prediction closely matches the ground truth (ray-tracing simulator) and outperforms other neural surrogates.
Notably, even evaluating the 2D path loss map (at a certain height), our 3D RayProNet shows a significant advantage over the 2D pipeline (PMNet). In our numerical experiments, the MSE score of PMNet is substantially lower than their 2D validations (approximately $\sim 10^{-2}$ in our isolated building environment and approximately $\sim 10^{-4}$ in their USC campus setting). The main reason for this difference is the size of the training dataset. The dataset in this experiment uses only 100 transmitters, whereas PMNet validates its pipeline with 19,016 configurations on the USC campus. Typically, a larger dataset size leads to better performance.
}

\begin{figure}[!ht]
\centering
\subfigure[GT]{
    \begin{minipage}[t]{0.22\textwidth}
        \includegraphics[width=\linewidth]{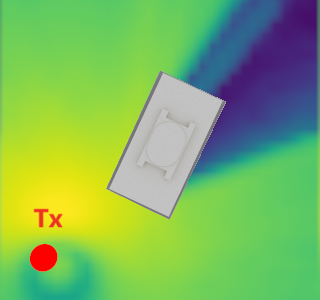}
        \label{fig: comparison with others: gt}
    \end{minipage}
}
\subfigure[Ours (3D)]{
    \begin{minipage}[t]{0.22\textwidth}
        \includegraphics[width=\linewidth]{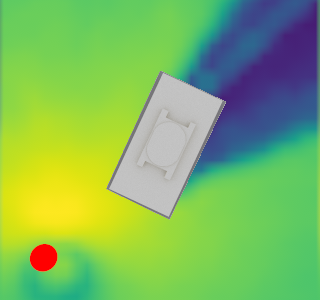}
        \label{fig: comparison with others: Ours}
    \end{minipage}
}
\\
\setcounter{subfigure}{0}	
\subfigure[MLP (3D)]{
    \begin{minipage}[t]{0.22\textwidth}
        \includegraphics[width=\linewidth]{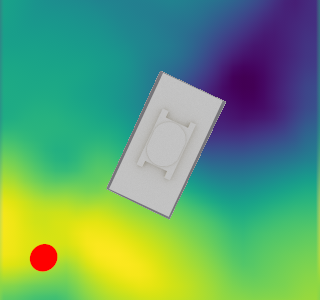}
        \label{fig: comparison with others: MLP}
    \end{minipage}
}
\subfigure[PMNet (2D)]{
    \begin{minipage}[t]{0.22\textwidth}
        \includegraphics[width=\linewidth]{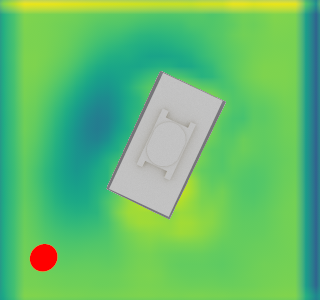}
        \label{fig: comparison with others: PMnet}
    \end{minipage}
}
\caption{{\textbf{Comparison with other neural surrogates}: Our method is compared to other neural surrogates, including PMNet and a standard MLP. The visualization illustrates our model's superior performance in predicting  the path loss map {at a certain height}.}}
\label{fig: Comparison with others}
\end{figure}

\begin{table}[h]
\begin{center}
\caption{\textbf{Comparison with other neural surrogates:} MSE loss and PSNR score comparison of power between ray-tracing results (ground truth) and various neural predictions (Ours, MLP, PMNet) in our isolated building environment. The upward arrow indicates better performance with larger values, while the downward arrow denotes better performance with smaller values. The best scores and lowest errors are highlighted in \textbf{bold} font.}
\label{Table: comparison with others}
\begin{tabular}{|c|c|c|c|c|}
 \hline
  - & Ours & MLP & PMNet\\
 \hline
 MSE $\downarrow$ & $\boldsymbol{3 \times 10^{-4}}$ & $4 \times 10^{-3}$ & $0.039$\\
 \hline
 PSNR $\uparrow$ & $\boldsymbol{35.24}$ & $23.90$ & $14.27$\\
 \hline
\end{tabular}
\end{center}
\end{table}

\subsubsection{\ZP{Verification through Ablation Experiment}}
{One of the key ingredients in RayProNet is the introduction of light probes. Hence, we will evaluate the impact of this module by performing an ablation experiment. If we remove the light probe module, receivers will directly shoot rays to find the $K$ closest point clouds, rather than the $n$ closest light probes. For this ablation experiment, we use a similar ray sampling strategy to \emph{NPLF} \cite{ost2022pointlightfields}. Both models are trained for 12 hours.
}

{The results of our ablation experiment are shown in Figure \ref{fig: Ablation results} and Table \ref{Table2Labe2}. These results align with our expectation that in both large outdoor scenes and small indoor scenes, it is common for a ray beam to be shot from an antenna but not reach any buildings (point clouds in our pipeline) nearby, causing the ray to be wasted as a default latent feature. The data in Figure \ref{fig: Ablation results} and Table \ref{Table2Labe2} support our analysis that power prediction is significantly limited when light probes are removed. Since light probes cover all areas in space, each antenna can always find a nearby light probe and extract propagation features from it. Hence, our proposed approach consistently aligns well with ray-tracing ground truth, even in large outdoor scenes.
}

\begin{figure}[!ht]
\centering
\subfigure{
    \rotatebox{90}{\scriptsize{~~~~~~~~~~~~GT}}
    \begin{minipage}[t]{0.21\textwidth}
        \includegraphics[width=\linewidth]{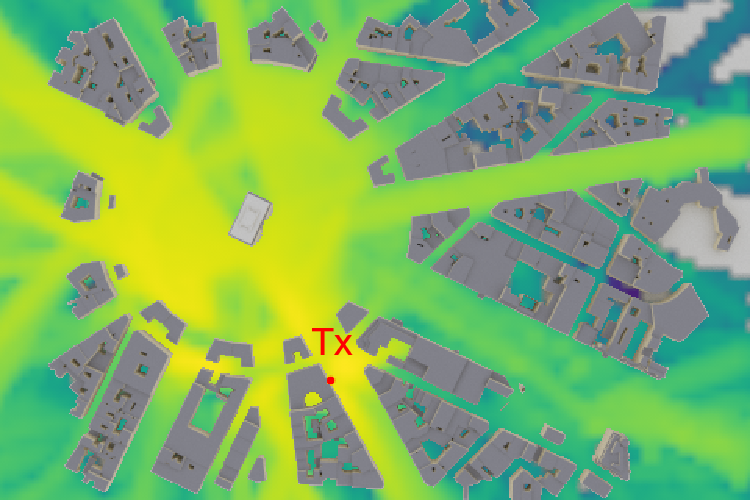}
    \end{minipage}
}
\subfigure{
    \begin{minipage}[t]{0.21\textwidth}
        \includegraphics[width=\linewidth]{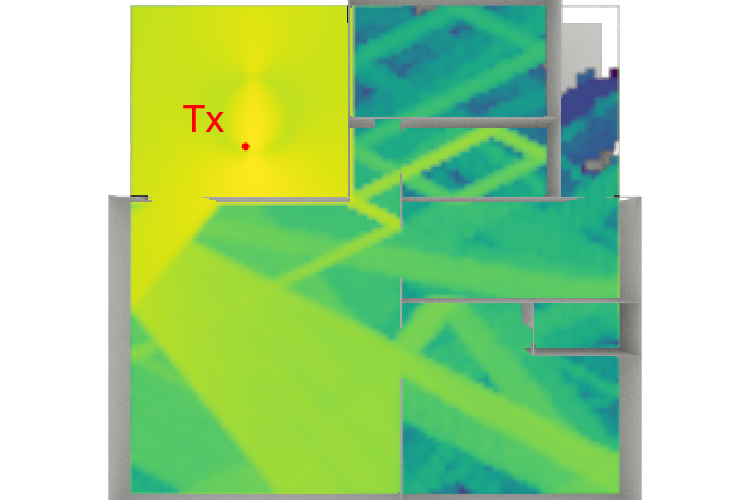}
    \end{minipage}
}
\\
\setcounter{subfigure}{0}
\subfigure{
    \rotatebox{90}{\scriptsize{~~~~~~~~~~~Ours}}
    \begin{minipage}[t]{0.21\textwidth}
        \includegraphics[width=\linewidth]{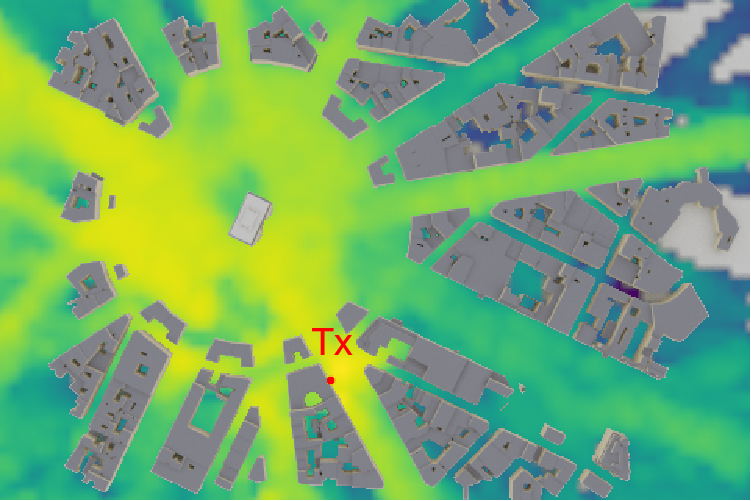}
    \end{minipage}
}
\subfigure[]{
    \begin{minipage}[t]{0.21\textwidth}
        \includegraphics[width=\linewidth]{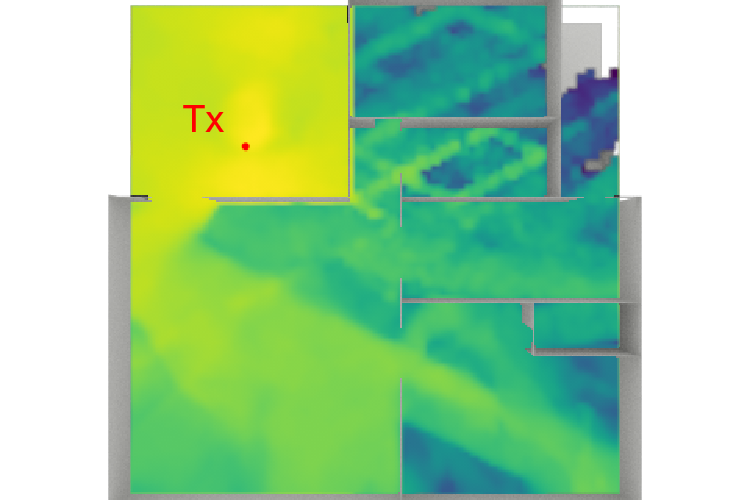}
    \end{minipage}
}
\setcounter{subfigure}{0}
\subfigure[]{
    \rotatebox{90}{\scriptsize{~~~~~w/o light probe}}
    \begin{minipage}[t]{0.21\textwidth}
        \includegraphics[width=\linewidth]{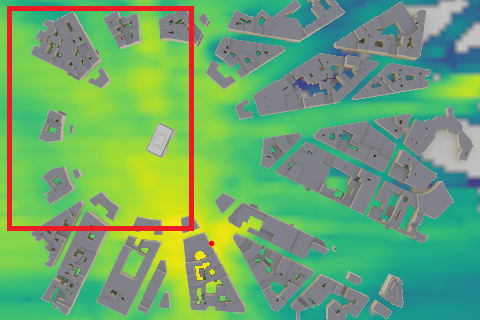}
        \label{subfig: ablation, etoile}
    \end{minipage}
}
\subfigure[]{
    \begin{minipage}[t]{0.21\textwidth}
        \includegraphics[width=\linewidth]{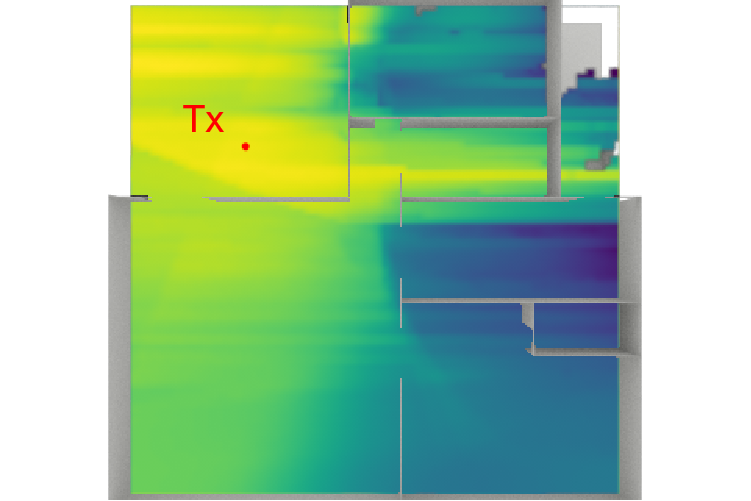}
        \label{subfig: ablation, wiindoor}
    \end{minipage}
}
\caption{\textbf{Ablation experiment:} This figure illustrates the validation of our proposed method by removing the light probe module from our approach. The validation is conducted in both large outdoor scenes (\emph{etoile}) and small indoor scenes (\emph{wiindoor}).}
\label{fig: Ablation results}
\end{figure}

\begin{table}[h]
\begin{center}
\caption{\textbf{Ablation experiment:} This table presents the MSE loss and PSNR score of power for both our dataset (\emph{etoile}) and \emph{WINERT}'s dataset (\emph{wiindoor}).}  \label{Table2Labe2}
\begin{minipage}{0.48\textwidth}

\begin{minipage}[t]{0.48\textwidth}
\makeatletter\def\@captype{table}
\begin{tabular}{|c|c|c|c|c|}
 \hline
  \emph{etoile} & Ours & Ablation\\
 \hline
 MSE $\downarrow$ & $\boldsymbol{0.0011}$ & $0.005$\\
 \hline
 PSNR $\uparrow$ & $\boldsymbol{29.38}$ & $23.07$\\
 \hline
\end{tabular}
\end{minipage}
\begin{minipage}[t]{0.48\textwidth}
\makeatletter\def\@captype{table}
\begin{tabular}{|c|c|c|c|c|}
 \hline
  \emph{wiindoor} & Ours & Ablation\\
 \hline
 MSE $\downarrow$ & $\boldsymbol{0.0017}$ & $0.022$\\
 \hline
 PSNR $\uparrow$ & $\boldsymbol{27.69}$ & $16.60$\\
 \hline
\end{tabular}
\end{minipage}

\end{minipage}

\end{center}
\end{table}

\subsection{\ZP{Evaluation in Large-scale, 3D Wireless Scenes}}

\subsubsection{Large-scale Environment}
\ZP{This subsection aims to validate the scalability of the proposed RayProNet in predicting EM propagation across different scene scales, ranging from small indoor rooms to expansive urban cities. 
By evaluating our model on three distinct scene scales, as depicted in Fig. \ref{fig: various scales}, we demonstrate its versatility and robustness. The results show a high degree of consistency with ray-tracing simulation results, confirming the accuracy of our model in diverse settings. 

In particular, the experiment involving a small-scale indoor room showcases our model's capability to accurately capture complex ray trajectories. Despite the inherent complexity of the ray paths, the model effectively \GC{recognizes} the intricate propagation patterns. This validation underscores our methodology's ability to handle a wide range of scenarios, making it suitable for applications in both indoor and outdoor wireless environments.

\GC{We also provide a time performance evaluation comparing our model to traditional ray-tracing (Table \ref{Table: Time performance}). The validation dataset consists of $25$ transmitters, $4$ antenna patterns, and $148 \times 121$ receivers ($100 \times 100$ in a small-scale indoor room scene). This setup results in 100 different configurations. The results show our methodology is at least 80 times faster than traditional ray-tracing, with an average time consumption of at most $3.2$ seconds per configuration.}
}

\begin{figure}[!ht]
\centering
\subfigure{
    \rotatebox{90}{\scriptsize{~~~~~~~~~indoor room}}
    \begin{minipage}[t]{0.21\textwidth}
        \includegraphics[width=\linewidth]{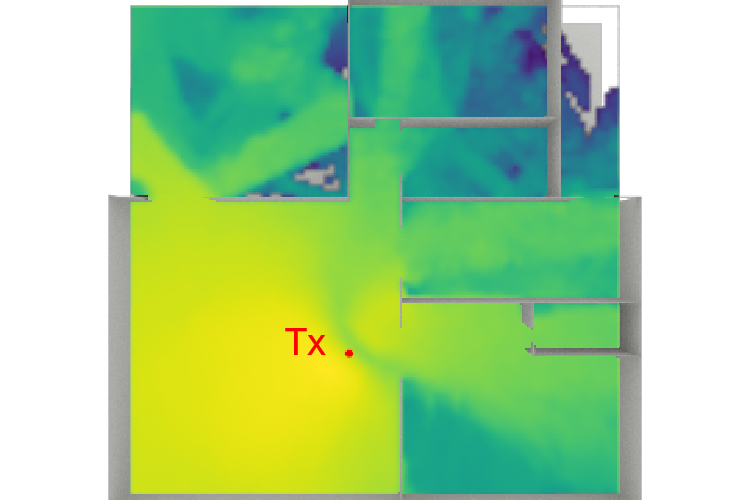}
    \end{minipage}
}
\subfigure{
    \begin{minipage}[t]{0.21\textwidth}
        \includegraphics[width=\linewidth]{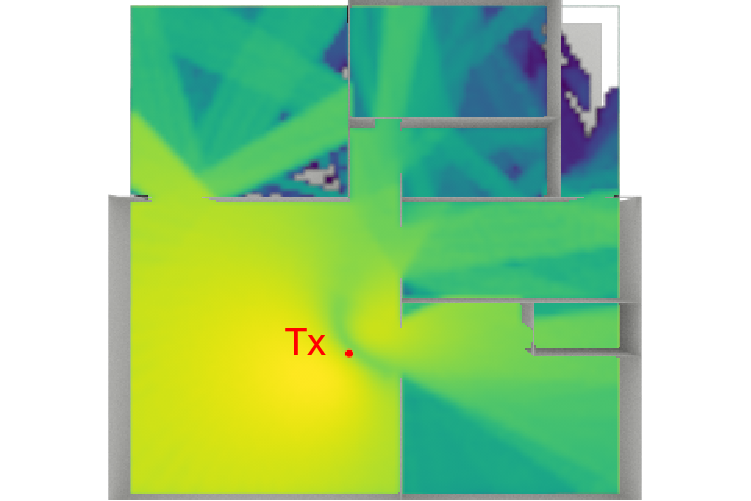}
    \end{minipage}
}
\\
\setcounter{subfigure}{0}
\subfigure{
    \rotatebox{90}{\scriptsize{~~~~~~~city blocks}}
    \begin{minipage}[t]{0.21\textwidth}
        \includegraphics[width=\linewidth]{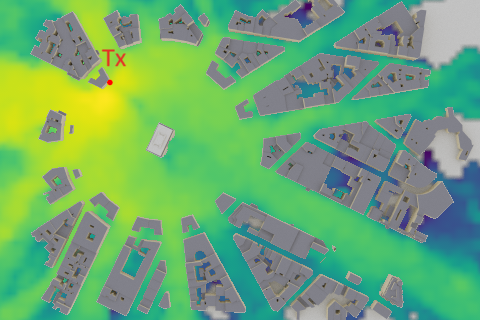}
    \end{minipage}
}
\subfigure{
    \begin{minipage}[t]{0.21\textwidth}
        \includegraphics[width=\linewidth]{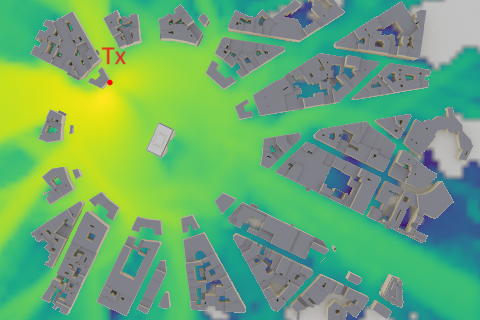}
    \end{minipage}
}
\\
\setcounter{subfigure}{0}
\subfigure[Ours]{
    \rotatebox{90}{\scriptsize{~~~~~~~~~urban city}}
    \begin{minipage}[t]{0.21\textwidth}
        \includegraphics[width=\linewidth]{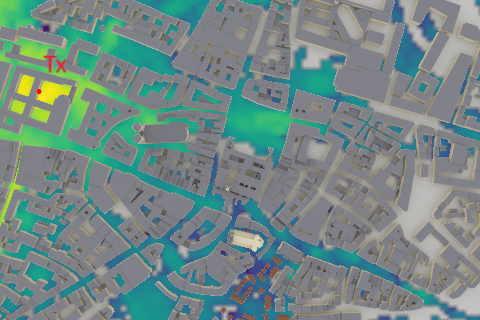}
    \end{minipage}
}
\subfigure[GT]{
    \begin{minipage}[t]{0.21\textwidth}
        \includegraphics[width=\linewidth]{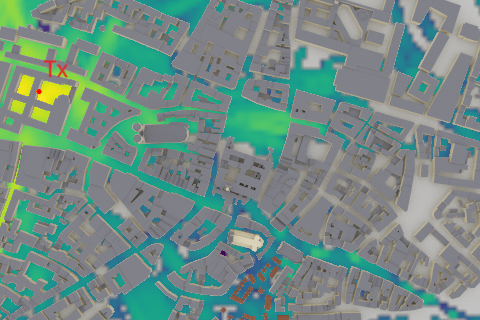}
    \end{minipage}
}
\caption{\textbf{Evaluation across various scales of environments}: Our pipeline supports a range of scene scales, from small indoor rooms to large urban cityscapes, demonstrating its flexibility and robustness.}
\label{fig: various scales}
\end{figure}

\begin{table}[h]
\begin{center}
\caption{\textbf{Runtime comparison between our model and ray-tracing:} In this table, we present a comparison of runtime performance between our model and ray-tracing with a validation set consisting of $25$ transmitters, $4$ antenna patterns, and $148 \times 121$ receivers ($100 \times 100$ in small-scale indoor room scene).}
\label{Table: Time performance}
\begin{tabular}{|c|c|c|c|c|}
 \hline
  Dataset & urban city & indoor room\\
 \hline
 Runtime (ours) & $\boldsymbol{32.86s}$ & $\boldsymbol{13.17s}$\\
 \hline
 Runtime (ray-tracing) & $2642s$ & $2771s$\\
 \hline
\end{tabular}
\end{center}
\end{table}

\subsubsection{Antenna Radiation Pattern}
\ZP{Furthermore, our RayProNet is capable of accommodating various types of \GC{trained} antenna radiation patterns as input. This versatility allows the model to adapt to different antenna configurations, enhancing its applicability in \GC{wireless planning scenarios}. The evaluation of these different antenna radiation patterns, as shown in Fig. \ref{fig: Evaluation Antenna radiation pattern}, reveals a substantial agreement between our predictions and the ray tracing results. \GC{Such capability is crucial for applications requiring detailed antenna placement, highlighting the practical utility and versatility of our proposed methodology in diverse wireless environments.}
}

\begin{figure}[!ht]
\centering
\subfigure{
    \rotatebox{90}{\scriptsize{~~~~~~~~~~~~Ours}}
    \begin{minipage}[t]{0.21\textwidth}
        \includegraphics[width=\linewidth]{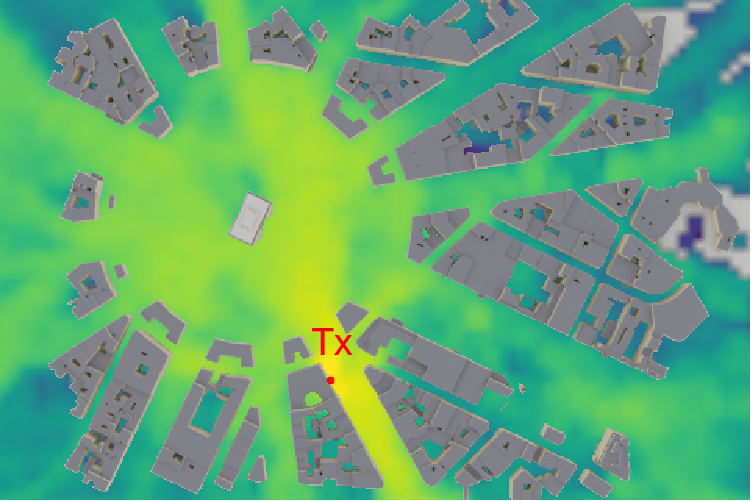}
    \end{minipage}
}
\subfigure{
    \begin{minipage}[t]{0.21\textwidth}
        \includegraphics[width=\linewidth]{figure/results/pred_cm_etoile_041_marked.png}
    \end{minipage}
}
\\
\setcounter{subfigure}{0}
\subfigure[]{
    \rotatebox{90}{\scriptsize{~~~~~~~~~~~~GT}}
    \begin{minipage}[t]{0.21\textwidth}
        \includegraphics[width=\linewidth]{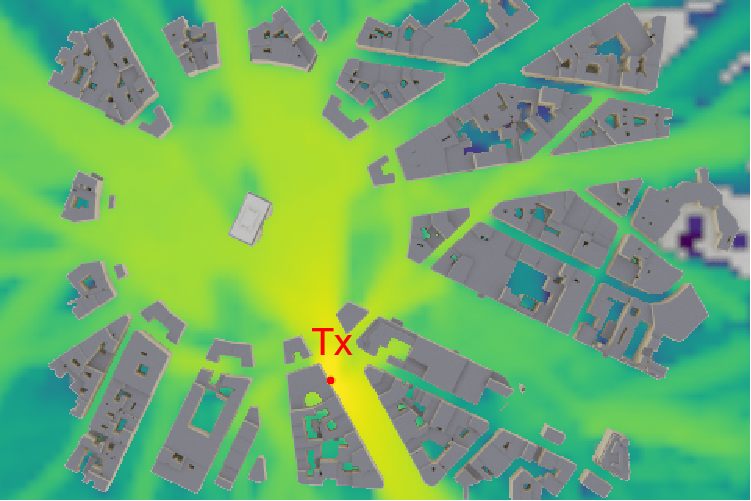}
        \label{fig: antenna 1}
    \end{minipage}
}
\subfigure[]{
    \begin{minipage}[t]{0.21\textwidth}
        \includegraphics[width=\linewidth]{figure/results/gt_cm_etoile_041_marked.png}
        \label{fig: antenna 2}
    \end{minipage}
}
\caption{\ZP{\textbf{Antenna radiation pattern}: Our pipeline support different antenna radiation pattern as input. In this figure, transmitters and receivers are equipped with \GC{isotropic (Fig. \ref{fig: antenna 1}) and patched antennas (Fig. \ref{fig: antenna 2}) respectively.}}}
\label{fig: Evaluation Antenna radiation pattern}
\end{figure}

\subsubsection{Quantitative Measurements}
\ZP{
So far, our results are primarily displayed in the format of 2D coverage maps for visualization. However, it is important to emphasize that our approach is essentially an end-to-end pipeline capable of predicting received signal strength at designated locations. To rigorously evaluate our model's performance, we selected five distinct receiver locations on the map: (-167.5 $\textrm{m}$, 22.5 $\textrm{m}$), (-162.5 $\textrm{m}$, 52.5 $\textrm{m}$), (-157.5 $\textrm{m}$, 62.5 $\textrm{m}$), (-147.5 $\textrm{m}$, 72.5 $\textrm{m}$), and (-137.5 $\textrm{m}$, 97.5 $\textrm{m}$). 

For each of these horizontal locations, we assessed the model's predictions at three different heights: 7.5 $\textrm{m}$, 10.5 $\textrm{m}$, and 13.5 $\textrm{m}$, resulting in a total of 15 evaluation points. This comprehensive selection allows us to test the model's accuracy and reliability across various spatial configurations. The precise locations of these points, along with the corresponding results, are illustrated in Figure \ref{fig: quantitative measurements}. This detailed analysis demonstrates our model's robustness and flexibility in accurately predicting power propagation in 3D environments.
}

\begin{figure}[!ht]
\centering
\subfigure[Receiver locations]{
    \begin{minipage}[t]{0.4\textwidth}
        \includegraphics[width=\linewidth]{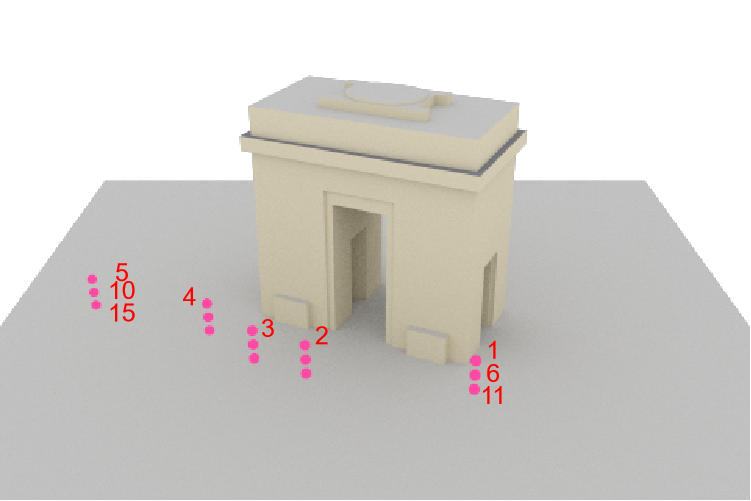}
        \label{fig: receiver points}
    \end{minipage}
}
\subfigure[Results]{
    \begin{minipage}[t]{0.4\textwidth}
        \includegraphics[width=\linewidth]{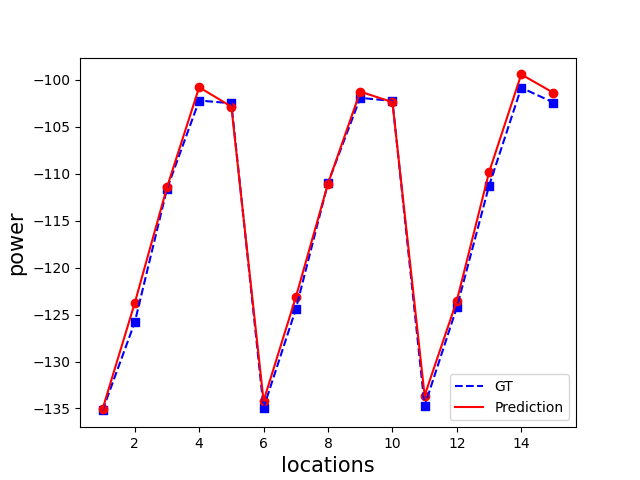}
        \label{fig: quantitative result}
    \end{minipage}
}
\caption{\ZP{\textbf{Quantitative measurements}: We selected 15 different receiver locations (Fig. \ref{fig: receiver points}) and quantitatively measure received power. The results are shown in Fig. \ref{fig: quantitative result}.}}
\label{fig: quantitative measurements}
\end{figure}
}

\section{Conclusion}
\ZP{To the best of our knowledge, this work represents the first effort in 3D neural wireless channel modeling capable of handling large-scale input scenes. Most prior works have focused on 2D image tasks that do not explicitly require explicit geometry representation. A recent work in \emph{Winert} \cite{WiNert2023} was primarily designed for small indoor scenes, as its pipeline necessitates mapping the intersection between a ray and a specific mesh triangle into a one-hot vector - an approach that is impractical for large scenes due to its excessive memory requirements.} 

\ZP{Our proposed method offers a significant advancement in rapid wireless channel modeling for extensive 3D scenes, achieving speeds \GC{$80 \sim 200$} times faster than GPU-accelerated ray tracing methods. This efficiency is particularly beneficial in scenarios where transmitter and receiver locations frequently change, such as in wireless deployment and planning.}

\ZP{Our framework does have a notable limitation: geometry and occlusion information are embedded within the neural networks. Consequently, any changes to the scene geometry necessitate re-training the pipeline. Future research will be focused on developing a more flexible framework capable of adapting to geometry changes without the need for re-training, enhancing its applicability and efficiency.}

\bibliographystyle{IEEEtran}
\bibliography{main}

\vfill

\end{document}